\documentclass[11pt]{article}

\usepackage{amssymb,latexsym}
\usepackage{epsfig}
\usepackage{eufrak}
\usepackage{amsmath}
\usepackage{mathrsfs}
\usepackage{color}

\newtheorem{theorem}{Theorem}
\newtheorem{lemma}{Lemma}

\newcommand{\be}{\begin{equation}}
\newcommand{\ee}{\end{equation}}
\newcommand{\bee}{\begin{eqnarray*}}
\newcommand{\eee}{\end{eqnarray*}}
\newcommand{\bel}{\begin{eqnarray}}
\newcommand{\eel}{\end{eqnarray}}
\newcommand{\bec}{\begin{cases}}
\newcommand{\eec}{\end{cases}}
\newcommand{\bem}{\begin{bmatrix}}
\newcommand{\eem}{\end{bmatrix}}

\newcommand{\li}{\left}
\newcommand{\ri}{\right}

\newcommand{\lc}{\lceil}
\newcommand{\rc}{\rceil}
\newcommand{\lf}{\lfloor}
\newcommand{\rf}{\rfloor}

\newcommand{\de}{\delta}

\newcommand{\se}{\theta}

\newcommand{\f}{\frac}
\newcommand{\sq}{\sqrt}
\newcommand{\cd}{\cdots}

\newcommand{\qu}{\quad}
\newcommand{\qqu}{\qquad}
\newcommand{\fa}{\forall}

\newcommand{\mscr}{\mathscr}
\newcommand{\mcal}{\mathcal}

\newcommand{\mrm}{\mathrm}

\newcommand{\sh}{\slash}

\newcommand{\iy}{\infty}

\newcommand{\pa}{\partial}

\newcommand{\bed}{\begin{description}}
\newcommand{\eed}{\end{description}}
\newcommand{\bei}{\begin{itemize}}
\newcommand{\eei}{\end{itemize}}
\newcommand{\ben}{\begin{enumerate}}
\newcommand{\een}{\end{enumerate}}
\newcommand{\bib}{\bibitem}
\newcommand{\beL}{\begin{lemma}}
\newcommand{\eeL}{\end{lemma}}
\newcommand{\beT}{\begin{theorem}}
\newcommand{\eeT}{\end{theorem}}
\newcommand{\sect}{\section}

\newcommand{\bpf}{\begin{pf}}
\newcommand{\epf}{\end{pf}}

\newcommand{\pfbox}{\hfill\mbox{$\Box$}}

\newenvironment{pf}{\paragraph*{Proof{\rm.}}}{\pfbox\bigskip}

\begin{document}

\title{{\bf Coverage Probability of Wald Interval for Binomial Parameters}
\thanks{The author is currently with Department of Electrical Engineering,
Louisiana State University at Baton Rouge, LA 70803, USA, and Department of
Electrical Engineering, Southern University and A\&M College, Baton
Rouge, LA 70813, USA; Email: chenxinjia@gmail.com}}

\author{Xinjia Chen}

\date{Submitted in April, 2008}

\maketitle

\begin{abstract}

In this paper, we develop an exact method for computing the minimum coverage probability of
Wald interval for estimation of binomial parameters.
Similar approach can be used for other type of confidence intervals.

\end{abstract}

\section{Wald Interval}

Let $X$ be a Bernoulli random variable with distribution $\Pr \{X = 1 \} = 1 - \Pr \{X = 0 \} = p \in (0, 1)$. Let $X_1, \cd, X_n$ be i.i.d.
random samples of $X$.  Let $K = \sum_{i = 1}^n X_i$.  The widely-used  Wald interval is $[L, U]$ with  lower confidence limit
\[
L =  \frac{K}{n}  - \mcal{Z}_{\de \sh 2} \sqrt{ \frac{ \frac{K}{n}
(1-\frac{K}{n}) } {n} } \]
 and upper
confidence limit
 \[
U = \frac{K}{n} + \mcal{Z}_{\de \sh 2} \sqrt{ \frac{ \frac{K}{n}
(1-\frac{K}{n}) } {n} }
\]
where $\mcal{Z}_{\de \sh 2}$ is the critical value such that
$\int_{\mcal{Z}_{\de \sh 2} }^\iy \f{1}{ \sq{2 \pi}  } e^{-
\f{x^2}{2}}= \frac{\delta}{2}$.  It has been discovered by Brown et
al. \cite {BCD2} that the coverage probability of Wald interval is
surprisingly poor.

\sect{Coverage Probability}

The coverage probability of Wald interval for binomial parameters has been investigated by \cite{BCD2} and other researchers by Edgeworth
expansion method and numerical methods based on discretizing the binomial parameter.  Here, we have obtained expression of the minimum coverage
probability of Wald interval, which requires only finite many evaluations of coverage probability.

\beT

Define
\[
\mrm{T}^{-} (p)  = \f{ 2 p + \se - \sq{ \se^2 + 4 \se p (1 - p) }
}{ 2 (1 + \se)  }, \qqu \mrm{T}^{+} (p)  = \f{ 2 p + \se + \sq{
\se^2 + 4 \se p (1 - p) }  }{ 2 (1 + \se)  }
\]
with $\se =  \f{\mcal{Z}_{\de \sh 2}^2}{n}$.  Define
\[
\mscr{L} (k) =  \frac{k}{n}  - \mcal{Z}_{\de \sh 2} \sqrt{ \frac{
\frac{k}{n} (1-\frac{k}{n}) } {n} }, \qqu \mscr{U} (k) = \frac{k}{n}
+ \mcal{Z}_{\de \sh 2} \sqrt{ \frac{ \frac{k}{n} (1-\frac{k}{n}) }
{n} }
\]
for $k = 0, 1, \cd, n$.  Define {\small \[ C_l (k) =  \Pr \{ \lc
\mrm{T}^{-} (\mscr{L}(k)) \rc \leq K \leq k - 1 \mid \mscr{L}(k) \},
\qu C_l^\prime (k) = \Pr \{ \lf \mrm{T}^{-} (\mscr{L}(k)) \rf + 1
\leq K \leq k - 1 \mid \mscr{L}(k) \}
\]}
for $k \in \{0, 1, \cd, n \}$ such that $0 < \mscr{L}(k) < 1$.
Define {\small \[ C_u (k) = \Pr \{ k + 1 \leq K \leq \lf \mrm{T}^{+}
( \mscr{U}(k) ) \rf \mid \mscr{U}(k) \}, \qu C_u^\prime (k) = \Pr \{
k + 1 \leq K \leq \lc \mrm{T}^{+} ( \mscr{U}(k) ) \rc - 1 \mid
\mscr{U}(k) \}
\]}
for $k \in \{0, 1, \cd, n \}$ such that $0 < \mscr{U}(k) < 1$.

Suppose $\se < 3$.  Then, the following statements hold true:

(I): $\inf_{p \in (0,1)} \Pr \{  L \leq p \leq U \mid p \}$ equals
to the minimum of {\small \[ \{ C_l (k): 0 \leq k \leq n; \; 0 <
\mscr{L}(k) < 1 \} \cup \{ C_u (k): 0 \leq k \leq n; \; 0 <
\mscr{U}(k) < 1 \}.
\]}

(II): $\min_{p \in (0,1)} \Pr \{  L < p < U \mid p \}$ equals to the
minimum of {\small \[ \{ C_l^\prime (k): 0 \leq k \leq n; \; 0 <
\mscr{L}(k) < 1 \} \cup \{ C_u^\prime (k): 0 \leq k \leq n; \; 0 <
\mscr{U}(k) < 1 \}.
\]}

\eeT

The proof of Theorem 1 is given in the next section.

\sect{Proof of Theorem 1}

We need some preliminary results.

\beL For $n > \f{\mcal{Z}_{\de \sh 2}^2}{3}$, both the lower and
upper confidence limits of Wald interval are monotonically
increasing with respect to $k$. \eeL

\bpf For simplicity of notation, let \[ z = \f{k}{n}.
\] Then, the upper confidence limit can be written as
\[
h (z) = z + \mcal{Z}_{\de \sh 2} \sq{z (1 - z) \sh n}.
\]
Similarly, the lower confidence limit can be written as
\[
g (z) = z - \mcal{Z}_{\de \sh 2} \sq{z (1 - z) \sh n}.
\]
To show that the upper confidence limit is monotonically increasing with respect to $k$,  it suffices to show that \[ \f{ \pa h(z) } { \pa z } >
0
\]
if $0 \leq h(z) \leq 1$.  Since
\[
\f{ \pa h(z) } { \pa z } = 1 +  \f{\sq{\se}}{2} \f{ 1 - 2 z }{ \sq{z (1 - z)}  },
\]
which is clearly positive for $0 < z \leq \f{1}{2}$, it remains to
show
\[
 \sq{\se} \f{ 1 - 2 z }{ \sq{z (1 - z)}  } > - 2, \qqu \fa z \in \li ( \f{1}{2}, 1 \ri )
\]
or equivalently,
\[
z(1 - z) > \f{\se}{4} (2 z - 1)^2, \qqu \fa z \in \li ( \f{1}{2}, 1 \ri ).
\]
Note that $h(z) < 1$ for $0 < z < \f{1}{1 + \se}$, and $h(z) > 1$ for $1 > z > \f{1}{1 + \se}$.

If $\f{1}{1 + \se} \leq \f{1}{2}$, i.e., $\se \geq 1$, then we are done, since $\f{ \pa h(z) } { \pa z } > 0$ for all $0 < z < \f{1}{1 + \se}
\leq \f{1}{2}$.  Otherwise, if $\se < 1$,  it suffices to show
\[
w(z) = z(1 - z) - \f{\se}{4} (2 z - 1)^2 > 0
\]
for $\f{1}{2} < z < \f{1}{1 + \se}$.  Since $z(1 - z)$ is decreasing
and $\f{\se}{4} (2 z - 1)^2$ is increasing for $1 > z > \f{1}{2}$,
$w(z)$ is decreasing for $\f{1}{2} < z < \f{1}{1 + \se}$. Therefore,
it suffices to have
\[
w \li ( \f{1}{1 + \se} \ri ) > 0,
\]
i.e.,
\[
\f{1}{1 + \se} \li (1 -  \f{1}{1 + \se} \ri ) - \f{\se}{4} \li ( \f{2}{1 + \se} - 1 \ri )^2 > 0,
\]
i.e.,
\[
\f{\se}{(1 + \se)^2 } - \f{\se}{4} \li ( \f{1- \se}{1 + \se} \ri )^2 > 0,
\]
i.e.,
\[
4 > (1 - \se)^2,
\]
which is guaranteed since $0 < \se < 3$.  This shows that the upper confidence limit is monotonically increasing with respect to $k$.  Observing
that $g(z) = 1 - h(1 -z)$, we have that the lower confidence limit is also monotonically increasing with respect to $k$.

\epf

Now we consider the minimum coverage probability.   By solving
equation
\[ \li ( p - \f{k}{n} \ri )^2 = \se \f{k}{n} \li ( 1 - \f{k}{n} \ri )
\]
with respect to $k$,  we can show that
\[
\Pr \{ L \leq p < U \mid \mscr{U}(k) \} =  \Pr \{ k < K \leq \mrm{T}^{+} (p) \mid \mscr{U}(k) \}, \qqu 0 < \mscr{U}(k) < 1
\]
\[
\Pr \{ L < p \leq U \mid \mscr{L}(k) \} =  \Pr \{ \mrm{T}^{-} (p) \leq K < k \mid \mscr{L}(k) \}, \qqu 0 < \mscr{L}(k) < 1
\]
\[
\Pr \{ L < p < U \mid \mscr{U}(k) \} =  \Pr \{ k < K < \mrm{T}^{+} (p) \mid \mscr{U}(k) \}, \qqu 0 < \mscr{U}(k) < 1
\]
\[
\Pr \{ L < p < U \mid \mscr{L}(k) \} =  \Pr \{ \mrm{T}^{-} (p) < K < k \mid \mscr{L}(k) \}, \qqu 0 < \mscr{L}(k) < 1
\]
Since both the lower and upper confidence limits of Wald interval
are monotone as asserted by Lemma 1, the proof of Theorem 1 can be
completed by making use of the above results and applying the theory
of coverage probability of random intervals established by Chen in
\cite{Chen}.

\end{document}